\documentclass[12pt]{article}
\usepackage{setspace}
\usepackage{epsfig,amsmath,amssymb} 
\usepackage[all]{xy}
\usepackage{ifpdf} 
\usepackage{graphicx}   
\input{Qcircuit}
\usepackage{amsfonts}
\usepackage{makeidx}  
\usepackage{epigraph}
\usepackage{etoolbox}
\makeatletter
\patchcmd{\epigraph}{\@epitext{#1}}{\itshape\@epitext{#1}}{}{}
\makeatother

\newcommand{\qed}{\hfill \mbox{\raggedright \rule{.07in}{.1in}}}

\renewcommand{\aa}{'}
\renewcommand{\ket}[1]{\left | #1 \right\rangle}

\newcommand{\qq}[1]{``#1"}

\begin{document}

\noindent \begin{center}{{\Huge{\bf{Constructor Theory of Life}}}}\\

\end{center}

\bigskip

\begin{center}
{{\bf{ Chiara Marletto}}} \\
\vspace{4pt}
 {Materials Department, University of Oxford}\\
\medskip
(November 2014) \\

\end{center}

\noindent

\bigskip\bigskip
\begin{center}
{\large{\bf{{Abstract}}}}
\end{center}

Neo-Darwinian evolutionary theory explains how the appearance of purposive design in the sophisticated adaptations of living organisms can have come about without their intentionally being designed. The explanation relies crucially on the possibility of certain {\sl physical processes}: mainly, {\sl gene replication} and {\sl natural selection}. 

In this paper I show that for those processes to be possible without the {\sl  design of biological adaptations} being encoded in the laws of physics, those laws must have certain other properties. The theory of what these properties are is not part of evolution theory proper, and has not been developed, yet without it the neo-Darwinian theory does not fully achieve its purpose of explaining the appearance of design.

To this end I apply Constructor Theory's new mode of explanation to provide an exact formulation of the appearance of design, of no-design laws, and of the logic of self-reproduction and natural selection, within fundamental physics.

I conclude that self-reproduction, replication and natural selection are possible under no-design laws, the only non-trivial condition being that they allow {\sl digital information} to be physically instantiated. This has an exact characterisation in the constructor theory of information. I also show that under no-design laws an accurate replicator {\sl requires} the existence of a \qq{vehicle} constituting, together with the replicator, a self-reproducer.

\section{Introduction}

Living entities display regularities unlike those observed in any other kind of matter in the universe. Although regular shapes of planets or crystals can be striking, these are explained by symmetries in the laws of physics; in contrast, even modest organisms, such as bacteria, display stupendously designed mechanisms, with {\sl many, different sub-parts coordinating to an overall function};  they perform transformations on physical systems with remarkable accuracy, retaining their ability to do so again and again - just as if they had literally been designed.  This {\sl appearance of design} was long considered evidence of intentional design \cite{PLATO, ARISTO, PAY}, and it does indeed require an explanation: Why is it there? How did it come into existence? 

The theory of evolution \cite{DAR} explains how the appearance of design can have been brought about by an undesigned physical process of variation and natural selection. It is a principle of the evolutionary theory that everything with the appearance of design must have come into existence by natural selection - directly (e.g. living organisms) or indirectly (objects that have literally been designed, such as cars or robots). 

In the modern neo-Darwinian synthesis \cite{DAWSELF, DAWEXT, DAWREVE}, the centrepiece of the explanation is a {\sl physical object} - the {\sl replicator} \cite{DAWSELF}: something that can be copied from generation to generation, by {\sl replication}, and {\sl selected} (between a set of variants) under the action of the environment. Instances of replicators in the earth's biosphere are \qq{genes}, i.e., portions of certain DNA molecule.\footnote{Subtleties about what portion of the genome is a gene,  what is selection, and at what level it occurs, are not relevant for the present discussion and will therefore be ignored.  It will suffice restricting attention to the logic of natural selection {\sl only}. }
Natural selection relies on gene replication, with occasional errors; the appearance of design is explained as adaptations for gene replication across generations; and the rest of the cell or organism (and sometimes other parts of the environment, e.g. nests, \cite{DAWEXT}) constitutes a {\sl vehicle} for the replicators.

Thus the neo-Darwinian theory of evolution relies on the laws of physics to permit replication and the processes essential to the latter -- including, as I shall explain, {\sl self-reproduction}. Therefore, for the theory to explain fully the appearance of design in the biosphere, it is essential that those processes be {\sl possible under laws of physics that do not contain the design of biological adaptations} - which I shall call {\sl no-design} laws.\footnote{For a discussion of how {no-design laws} differ from generically fine-tuned laws, or {\sl bio-friendly} laws, as in \cite{DAV05}, see section \ref{SEC:SR}.} 

In this paper I show that those physical processes are indeed possible, provided that those laws have certain other properties. Although the theory of what these properties are does not belong to evolutionary theory proper, the neo-Darwinian theory cannot fully explain the appearance of design without it.\footnote{The term \qq{life} in the title is just a shorthand to refer to all those processes that the theory of evolution relies upon to explain \qq{life} on earth. (It will not be necessary here to define \qq{life} more precisely than that.)} 

To explain why, we need to examine more precisely the physical processes central to the theory of evolution. A{ replicator} is an object $R$ that is capable of undergoing {replication} (i.e., being copied), as in this schematic pattern:
\begin{figure}[h]
\centering
\includegraphics[scale=0.2]{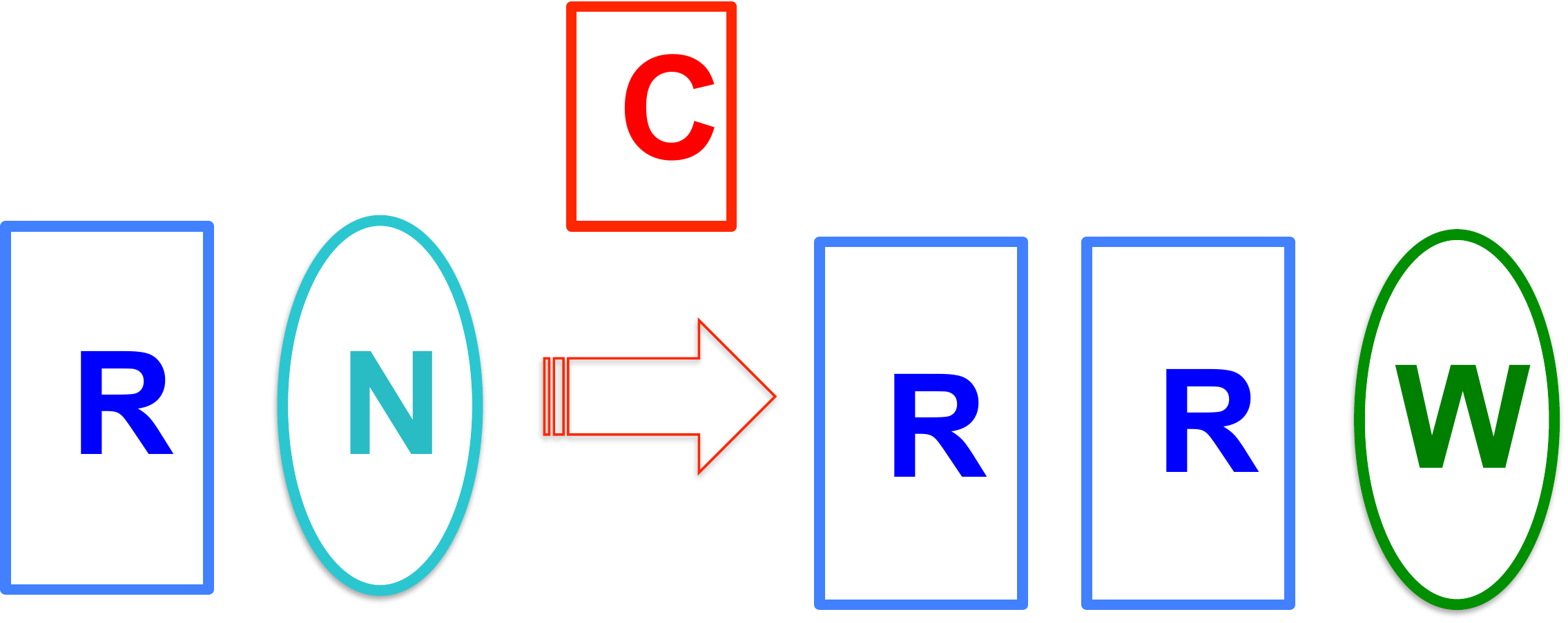} 

\end{figure}


${C}$ is the copier, acting on some raw material $N$ (possibly producing waste products $W$); in living entities the copier is included in the cell - whose self-reproduction is thus essential to gene replication over many generations. 

A {\sl self-reproducer} is an object $S$ capable of undergoing the physical transformation

%

\begin{figure}[h]
\centering
\includegraphics[scale=0.2]{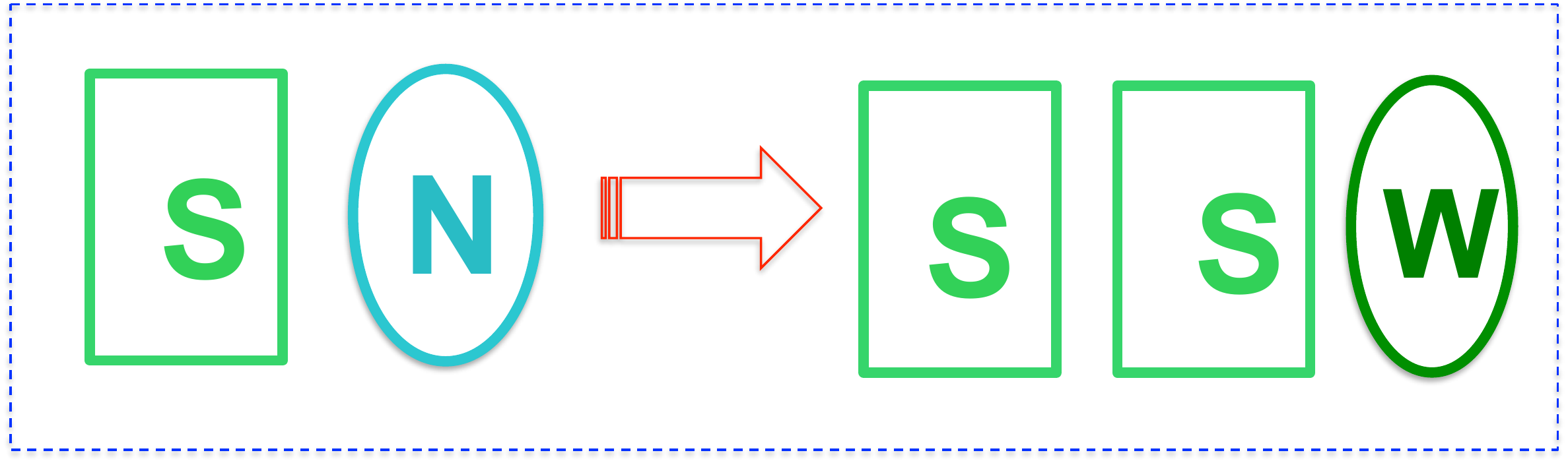} 
\end{figure}

where the raw materials $N$ need not contain the means to create another $S$, and the whole system could be isolated. For present purposes it suffices to model the {logic} of self-reproduction only as it occurs in early life and pre-life - for instance, sexual reproduction need not be modelled and the environment can be assumed to be non-biological. 

In that context the difference between a replicator and a self-reproducer is that the latter does not rely on any mechanism other than itself and the laws of physics to cause the construction of the new instance, whereas a replicator depends on a copying mechanism outside itself. As I shall explain, under no-design laws this implies that an accurate self-reproducer consists of a \qq{vehicle} and a replicator, and its self-reproduction occurs by copying the replicator and re-constructing the vehicle afresh. This \qq{replicator-vehicle} logic (see section \ref{SEC:SR}) was discovered by von Neumann, \cite{VON}, and its relevance in biology thoroughly analysed in \cite{DAWREVE}, \cite{DYS}.

In the biosphere self-reproduction is approximated to various accuracies. There are many poor approximations to self-reproducers - e.g., crude replicators such as crystals, short RNA strands and autocatalytic cycles involved in the {origin of life} \cite{MAZA}. Being so inaccurate, they do not require any further explanation under no-design laws: they do not have appearance of design, any more than simple inorganic catalysts do.\footnote{ The very existence of catalysts might be a sign of fine-tuning in the laws of physics, but not fine-tuning {\sl for biological adaptations}, with which we are concerned here. }

In contrast, actual gene-replication is an impressively accurate physical transformation, albeit imperfect. But even more striking is that living cells can self-reproduce to high accuracy in a variety of environments, reconstructing the vehicle afresh, under the control of the genes, in all the intricate details necessary for gene replication. 
This is {\sl prima facie} problematic under no-design laws: how can those processes be so accurate, without their design being encoded in the laws of physics? This is why some physicists - notably, Wigner and Bohm, \cite{WIG}, \cite{BOH} -  have even claimed that {\sl accurate} self-reproduction of an organism with the appearance of design requires the laws of motion to be \qq{tailored} for the purpose -- i.e., they must contain its design \cite{WIG}. 

These claims, stemming from the tradition of incredulity that living entities can be scientifically explained, \cite{NAG}, highlight a problem. The theory of evolution must be supplemented by a theory that those physical processes upon which it relies are provably compatible with no-design laws of physics. No such theory has been proposed; and those claims have not been properly refuted.

Indeed, the central problem here -- i.e., {\sl whether and under what circumstances accurate self-reproduction and replication are compatible with no-design laws} -- is awkward to formulate in the {\sl prevailing conception of fundamental physics}, which expresses everything in terms of predictions given some initial conditions and laws of motion. 

This mode of explanation can only approximately  express emergent notions such as the appearance of design, no-design laws, etc. 

Von Neumann, who attempted to investigate self-reproduction within this framework,  got as far as discovering its essential (replicator-vehicle) logic, \cite{VON}. However his use of the prevailing conception forced his analysis to be in terms of predictions: thus he attempted without success to provide the design of an actual self-reproducer in terms of atoms and microscopic interaction.  He finally produced a viable toy model, \cite{VONBU}, within cellular automata, but at the cost of severing the connections with actual physics. That model is thus inadequate to address the current problem - whether self-reproduction is compatible {\sl with the actual laws of physics} un-augmented by any design of adaptations.  

The prevailing conception also forces a misleading formulation of the problem, as: what initial conditions and laws of motion {\sl must} (or must probably) produce accurate replicators and self-reproducers (with some probability)? 
But what is disputed is whether such entities are {\sl possible} under no-design laws. 

More generally, it cannot express the very explanation provided by evolutionary theory -- that living organisms {\sl can} have come about without intentionally being designed. It would have aimed at proving that they {\sl must} occur, given certain initial conditions and dynamical laws.

To overcome these problems I resort to a newly proposed theory of physics, constructor theory. \cite{DEU, DEUMA, NEWSCI}. 
It provides a new mode of explanation, expressing all laws as statements about which transformations are {\sl possible}, which are {\sl impossible} and why. 

This brings {\sl counterfactual statements} into fundamental physics, which is key to the solution. The explanation provided by the theory of evolution is already constructor-theoretic: it is {\sl possible} that the appearance of design has been brought about without intentionally being designed; so is our problem: are the physical processes essential to the theory of evolution - i.e., self-reproduction, replication and natural selection - {\sl possible} under {\sl no-design} laws? 

I shall show that they are (in section 2-3) provided that those laws of physics allow the existence of media that can instantiate (digital) {\sl information} (plus enough time and energy). {\sl Information} has an exact physical characterisation in the constructor theory of information \cite{DEUMA}. 

I also show that under no-design laws an accurate self-reproducer {\sl requires} an accurate (i.e., {\sl high-fidelity}) replicator, and {\sl vice versa}. Thus, the replicator-vehicle logic von Neumann envisaged is here shown to be {\sl necessary} for accurate self-reproduction to be possible under such laws. This provides physical foundations for the relation between \qq{metabolism} and replication (as defined by Dyson, \cite{DYS}). In addition, that vehicles are necessary to high-quality replicators under our laws of physics (despite replicators being the conceptual pillar of evolutionary theory), informs the current debate about the necessity of organisms. The latter was recently doubted by Dawkins \cite{DAWPAR}: \qq{ Just as life did not have to become multicellular [...] so living materials did not have to become packaged into discrete, individual organisms [..] behaving as unitary, purposeful agents. The only thing that is really fundamental to Darwinian life is self-replicating, coded information - genes, in the terminology of life on this planet.}. 

Constructor Theory's mode of explanation also delivers an exact physical expression  of the notions of  the appearance of design, no-design laws, and of the {\sl logic} of self-reproduction and natural selection.\footnote{The model is intended to be faithful {\sl only} insofar the {\sl logic} is concerned. Most realistic details of these processes are irrelevant to their logic, so they shall be neglected.} 

Finally, Wigner's argument implies that accurate self-reproduction is incompatible particularly with {\sl quantum theory}, thus challenging its universality - a claim that others, with different motivations, have also made \cite{BRA, CHAPRA, PATI}. I shall demonstrate (in section 4) a quantum-mechanical (kinematical) model of the {\sl logic} of self-reproduction, updating von Neumann\aa s, thus rebutting those claims.  This, incidentally, clarifies how self-reproduction differs from cloning a quantum state (which has occasionally caused some confusion \cite{BRA}). It also vindicates that self-reproduction - and even (possibly artificial) self-reproducers employing quantum coherence - are compatible with quantum theory.

\section{The problem}

I shall introduce constructor theory's tools to re-formulate the problem in constructor-theoretic terms. 

Constructor theory is a new fundamental theory of physics. First, it provides a paradigm to express the other laws of physics, to be expressed {\sl solely} as statements about which transformations are {\sl possible}, which are {\sl impossible} and why. Guesses at those laws - e.g.,  current physical theories such as general relativity and quantum mechanics - it calls {\sl subsidiary theories}. In addition, it also proposes new laws - principles - constraining the subsidiary theories. Here it suffices to know that the principles are obeyed by all known laws of physics, nor do they themselves contain the design of biological adaptations (see \cite{DEU}, \cite{DEUMA}).

The properties of a physical system ${\bf M}$ are {\sl attributes}, defined as sets of states of ${\bf M}$. We say that ${\bf M}$ (say, a collection of atoms) has the attribute $X$ (say, being a car, or a self-reproducer) if it is in any of the states in $X$.

Constructor theory's main elements are {\bf tasks}. 
A task $T$ is the abstract specification of a transformation
$$
T=\left \{ x_1\rightarrow y_1, x_2\rightarrow y_2,\dots\;, x_n\rightarrow y_n  \right \}
$$
as a {\sl set of input/output pairs} of attributes $\{x_i\}$,  $\{y_i\}$ of the {\bf substrates} (the physical systems being transformed).

Tasks form an algebra under parallel and serial composition, and are composable into networks to form other tasks \cite{DEUMA}.

A physical system with some attribute ${\mathfrak {C}}$ is {\bf a constructor}, {\bf capable of performing the task} $T$ if: 

\begin{itemize}

\item{} whenever presented with the substrates with any of the legitimate input attributes of $T$, ${\mathfrak {C}}$ delivers it with the corresponding output attribute, as follows: 

 \begin{center}
Input attributes of substrates $\overset{{\rm {{\mathfrak C}}}}{\implies}$ Output attributes of substrates,
\end{center}
where ${\mathfrak {C}}$ and the substrates {\sl jointly constitute an isolated system};
\item{} ${\mathfrak {C}}$ retains the ability to do so again. 
\end{itemize}

A task is {\bf impossible} if it is forbidden by the laws of physics (e.g., building a perpetual motion machine); otherwise, it is {\bf possible}. 

Under our laws of physics, only approximate constructors exist, e.g. catalysts or robots. They have non-zero error rates and deteriorate with use. Hence, that a task is {possible} means that the laws of physics impose no limit, short of perfection, on how accurately it could be performed, nor on how well objects capable of approximately performing it could retain their ability to do so. The term \qq{constructor} is a placeholder for the (infinite) sequence of approximations to the ideal behaviour of a constructor. 

Both replication and self-reproduction find an exact expression in constructor theory. The {\bf replication} of a set $\Sigma$ of attributes of a system is the task
\begin{equation}\bigcup\limits_{X\in \Sigma}{\left\{ \left( X,N \right)\to \left( X,X, W \right) \right\}}\;\label{RR}\end{equation}
on the composite system ${\bf M_1}\oplus {\bf M_2}$ of the {\sl source} substrate (containing the attribute to be copied) and the {\sl target} substrates  (onto which the attribute is copied). $X\in \Sigma$ is an attribute of ${\bf M_1}$, being replicated; $N$ some receptive attribute of ${\bf M_2}$ and $(X, W)$ the output attribute, including waste products $W$.  For example, $\Sigma$ could be a set of musical notes; or the set of alleles (variants of a gene) or the set of nucleotides. \footnote{It is a set of objects that can be copied - not necessarily containing a set of control instructions.}

{\bf Self-reproduction} is a {\sl construction} where the self-reproducer $S$ is a constructor, for constructing another instance of itself given raw materials $N$ containing neither $S$ nor constructors for $S$:
\begin{equation}
N\overset{\rm S}{\implies} (S,W) \;\label{REP2}
\end{equation} allowing for waste products. $S$ is the specification of all properties necessary for \eqref{REP2} to occur, given the laws of physics. 

 {\bf No-design laws} can be expressed exactly in constructor theory, too.

First, I define {\qq{{\bf generic resources}} as substrates that exist in {\sl effectively unlimited} numbers. In the context of early life on this planet, these include only elementary entities such as photons, water, simple catalysts and small organic molecules.


It has sometimes been proposed that the very existence of laws of nature constitutes a form of \qq{design} in them, \cite{CART}.  In contrast, for present purposes {no-design laws} are those that do not contain the design of biological adaptations - i.e., of what the theory of evolution aims at explaining: for the problem here is whether the physical processes assumed by the theory of evolution are possible under such laws. 

Consequently I require {no-design laws} to satisfy these conditions: 

\begin{itemize}

\item{} Generic resources can only perform {\sl a few} tasks, only to a finite accuracy, called {\sl elementary tasks}. These are physically simple and contain no design (of biological adaptations).  Familiar examples are spontaneous, approximately self-correcting chemical reactions, such as molecules \qq{snapping} into a catalysts regardless of any original small mismatch. 
\item{} No good approximation to a constructor for tasks that are non-elementary can ever be produced by generic resources acting on generic resources {\sl only}.  
\end{itemize}

Under no-design laws, the generic resources and the interactions available in nature are allowed to contain only those approximate constructors that unequivocally do not have the design of those very adaptations the theory of evolution is required to explain.\footnote{Thus, quantum theory is no-design in this sense: the initial state of the universe is a uniform, low entropy state; the elementary interactions allowed in quantum theory do not contain the design of biological adaptations.} Examples of laws that would violate these conditions are: laws including accurate constructors, such as bacteria, in the generic resources; laws with \qq{copy-like} interactions, designed to copy the configuration of atoms of a bacterium onto generic resources; laws permitting spontaneous generation of a bacterium {\sl directly} from generic resources only; laws permitting only mutations that are systematically directed to {\sl improvements} in a certain environment.

The exact characterisation of no-design laws is a departure from the prevailing conception - which can at most characterise them as being typical, according to some measure, in the space of all laws. The latter is unsuitable for present purposes, as the choice of the measure is highly arbitrary. Moreover, it is misleading: some laws that may be untypical under some natural measure - such as the actual laws of physics, because of, say, local interactions -  need not contain the design of biological adaptations, thus qualifying as no-design in this context. Furthermore, laws with the design of biological adaptations are a proper subset of those laws that in the context of {anthropic} fine tuning have been called \qq{bio-friendly}: those having features - such as local interactions, or special values of the fine-structure constant, etc. - which, if slightly changed, would cause life as we know it to be impossible. These features, though necessary to life, are {\sl not specific to life}: their variation would make impossible many other phenomena, non specifically related to biological adaptations.

The problem can now be restated in constructor theory, as: {\sl are accurate self-reproducers and replicators possible under no-design laws?} 

I shall prove that an accurate self-reproducer is possible under no-design laws, provided they allow information to be physically instantiated; from this it will follow that an accurate replicator is possible too, provided that it be contained in a self-reproducer, (sections \ref{SEC:CAR} - \ref{NAT}).

I will assume that the raw materials of self-reproduction ($N$ in \eqref{RR}, \eqref{REP2}) comprises generic substrates {\sl only}. This over-stringent assumption rules out the realistic situation that they contain other organisms; but it is acceptable for present purposes because if accurate self-reproduction and replication are allowed under these over-stringent requirements, so are they when the generic resources contain also living organisms.

Before presenting the argument, I shall recall the basics of the constructor theory of information (section \ref{SEC:INF}). This is crucial to give an exact characterisation of what it means for the laws of physics to allow information to be physically instantiated.

\subsection{Information}\label{SEC:INF}

Replication when regarded as copying is intimately connected with  {\sl information}. This has inspired some information-based approaches to fundamental problems in biology, \cite{MAYINFO}. Until recently, information had no place in fundamental physics: expressions such as \qq{information being instantiated in a physical system} were inherently approximate, or fuzzy. But the {\sl constructor theory of information} has now incorporated information within fundamental physics, \cite{DEUMA}, providing an exact, physical characterisation to those expressions, as follows.

A set of attributes $\Sigma$ is an {\bf information variable} \cite{DEUMA} if the task of performing any {\sl permutation} over $\Sigma$ (allowing for waste), and the {\sl replication task} over $\Sigma$, as in \eqref{RR}, are all possible.
The attributes of an information variables are called {\bf information attributes}. An {\bf information medium} is a substrate some of whose attributes constitute an information variable. 

Information media must obey the {\sl interoperability principle} \cite{DEUMA}: the composite system of two information media with information variables $\Sigma_1$ and $\Sigma_2$, is an information medium with information variable $\Sigma_1\times \Sigma_2$. This is a physical principle: it requires there to be interactions such that information is \qq{copiable} from one information medium to any other. 

Thus, whether or not information media exist, i.e., whether or not information can be instantiated in physical systems, depends on the laws of physics. The intuition about replication being central to information is now expressed as a physical law: laws of physics permitting information media must allow information variables - i.e., replicable sets of attributes as in \eqref{RR}. 

A physical system ${\bf M}$ {\sl instantiates information} if it is an information medium in one of its information attributes (belonging to some information variable $\Sigma$) and that the task of giving it any other attribute in $\Sigma$ (allowing for waste) is possible. This is an exact physical requirement: for this to be possible, certain interactions must be available in nature. It is also an intrinsic, {\sl counterfactual}, property of ${\bf M}$.

A constructor $C$ for the replication task on some information variable $\Sigma$ 

\begin{equation}
(X, N)\overset{ C} {\implies}(X, X,W) \;,\;\;\forall X\in\Sigma\nonumber
\end{equation}
is called a {\bf copier} of $\Sigma$.

Of its substrates, one  - the {\sl target}  - is changed from having the attribute $N$ to having the attribute $(X,W)$; the other - the {\sl source}, initially having one of the attributes $X$ in $\Sigma$, to be replicated - remains {\sl unchanged} (but it may change temporarily during the copy process).

Therefore (by definition of a constructor) $C$ and the source substrate with the attribute $X$ constitute a constructor $C[X]$ performing the task $T_X=\{N\rightarrow (X,W)\}$ on the target. 
The information attribute $X$ in the source acts as a constructor, {\sl instructing} $C$ to perform the task $T_X$ on $N$. See the figure \ref{PROG}.
%
\begin{figure}[h]\label{PROG}
\centering
\includegraphics[scale=0.3]{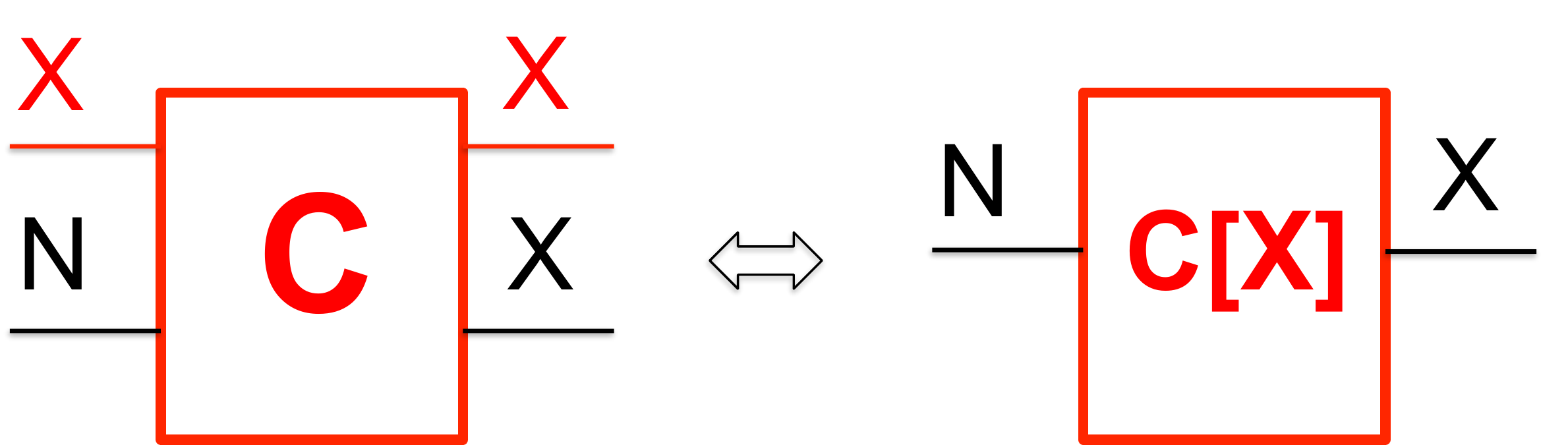} 
\caption{\small Two equivalent representations of a copier $C$ (waste $W$ omitted). On the left, $C$ is a constructor with two substrates (represented by lines): the source, that remains unchanged; and the target, that is changed.  On the right, $C$ and the source substrate with the attributes $X$ constitute the constructor $C[X]$ performing the task $T_X=\{N\rightarrow X\}$ on the target, for all $X \in \Sigma$. The copier is a simple example of a programmable constructor.}
\end{figure}

In general, a {\bf programmable constructor} is a constructor $V$ that has, among its input substrates, an information medium ${\bf M}$ that can have any of the attributes $P$ in an information variable, with the property that ${\bf M}$ with any of those information attributes is itself a constructor. The information instantiated in ${\bf M}$ is an {\sl abstract constructor} - an instance of \qq{information with causal power}, \cite{WADA}. $V[P]$ is a constructor for the task $T_P$, $P$ is the program for the task $T_P$ and $T_P$ is in the repertoire of $V$. For example, $V$ could be the ribosome, $P$ the sequence that, when inserted in $V$, would cause $V$ to perform the task $T_P$ of constructing a particular polypeptide chain. 

\section{The theory of evolution by natural selection is compatible with no-design laws}

I shall now show that under no-design laws accurate self-reproducers and accurate replicators are not forbidden, provided only that the laws permit the existence of information media (and enough resources). This will vindicate that the theory of evolution by natural selection is compatible with those laws (and thus, in particular, with the current theories of physics). 

My argument includes three steps. First I establish that an accurate constructor for a generic task is compatible with no-design laws (section \ref{SEC:CAR}), provided that it contains a replicator, instantiating a recipe for that task. As a special case, I show that accurate self-reproducers are compatible with no-design laws (section \ref{SEC:SR}), provided that they implement the \qq{replicator-vehicle} logic; it follows that so are accurate replicators, and that they require there to be a self-reproducer. Finally I show that the logic of natural selection is compatible with no-design laws (section \ref{NAT}).

\subsection{An accurate constructor must contain a replicator}\label{SEC:CAR}

A task $T$ being possible means that for any given accuracy (short of perfection) the laws of physics permit an approximate constructor capable of performing the task to that accuracy.

Consider a possible, non elementary task $T$ and an object $F$ that can perform $T$ to a high accuracy \footnote{It is the subsidiary theory that provides specific measures of accuracy.} $\epsilon$. For instance, $T$ could be the task of constructing a car from generic substrates and $F$ a generalised car factory, including all the processes converting raw materials such as iron, etc., into a car. 

The approximate constructor $F$ {\sl executes} a procedure - a {\bf recipe} - to perform the task $T$ to accuracy $\epsilon$. I will show that $F$ must include a replicator and a programmable constructor; and that the recipe must have a hierarchical structure and be instantiated in the replicator. 

No-design laws contain no good constructor for $T$, such as $F$ - neither in the elementary interactions, nor in the generic resources. 
Hence the {\sl recipe} used by $F$ to perform $T$ must be decomposable into steps (not necessarily sequential) that {\sl are} allowed by no-design laws. That is to say, sub-recipes - procedures to perform sub-tasks that are executed by sub-constructors contained in $F$. To avoid infinite regress, two conditions must be fulfilled.

One is that the subtasks be {\sl non-specific} to $T$. For instance, when T is the task of constructing a car, the subtasks are those of constructing sub-parts of the car - e.g., door handles, windows, etc. 
Hence, the constructor $F$ must include two parts: One -- which I call $V$ -- performs $T$ {\sl blindly}, i.e., subtask by subtask, and it is non-specific to $T$, because so are the subtasks. The rest of $F$ -- which I call $P$ -- is specific to $T$ and instantiates the recipe for $T$: it specifies the sequence of the subtasks, thus controlling $V$. Hence $F$ can be described as a programmable constructor, $V$, programmed with a program $P$ having {\sl the same logic as the recipe}: it has a modular structure $P=(p_1, p_2, \cdots, p_N)$ where each instruction $p_i$ takes values in an information variable and tells $V$ which sub-task to perform, when, on the substrates\footnote{This is a schematic representation: $P$ need not have a linear structure.}.  $V$ is non-specific to $T$ because it must also be capable of executing other programs - different combinations of the elementary units $p_i$. For example, a car factory contains robots executing sub-recipes to construct the car's doors. These robots contain sub-robots to construct handles, windows, etc., which could be used to construct other objects than cars. 

The other condition is obtained by applying the same reasoning recursively to the subtasks. If they, too, are non-elementary, they require a recipe that is decomposable into non-specific sub-recipes. The base for the recursion - for $T$ to be performable to that particular accuracy - is provided by the {\sl elementary sub-recipes} of the recipe for $T$ being elementary tasks - which can be performed by (approximations to) constructors that are available in nature, as generic resources. 

Note that these elementary sub-tasks need not be specified in the recipe: they are implicit in the laws of physics. For instance, the elementary steps in the car recipe are tasks like, say,  \qq{oxidise the aluminium coating}, and occur simply by leaving the substrate exposed to air. 


Under no-design laws, any (approximation to a) constructor {\sl wears out} after a finite time. Therefore $F$, to perform the task $T$ to the accuracy $\epsilon$, must undergo a process of {\bf maintenance}, defined as one whereby a new instance of $F$ - i.e., of $P$ and $V$ - is brought about, from generic materials, before the former one stops working. In the case of the car factory, this is achieved by replacing old subparts of the robots, assembly lines, etc. and by {preserving the programs} they run.

To avoid an infinite regress, implementing the maintenance must not in turn require the recipe $P$ for $T$. Also, the design of the recipe $P$ cannot be in the laws of physics. Thus, the only other possibility is that the new instance of $P$ is brought about by {\sl blind replication} of the recipe $P$ contained in the former instance - i.e., by replicating its subunits $p_i$ (that are non-specific to $T$). We conclude that, under no-design laws, the substrate instantiating the recipe is necessarily a {\bf modular replicator}: a physical object that can be copied blindly, an elementary subunit at a time. In contrast, $V$ - the non-specific component of $F$ - is {\sl constructed} anew from generic resources. 

Moreover, under no-design laws {\sl errors} can occur: thus, to achieve high and improvable accuracy, the recipe must include {\bf error-correction}. In the car factory, this includes, say, controlling the functionalities of the subcomponents (e.g., fine checks on the position of doors, wheels, etc.). Hence {\sl the recipe $P$ must contain information about the task $T$, informing the criterion for error detection and correction}.

The information in the recipe is an abstract constructor that I shall call {\bf knowledge} (without a knowing subject \cite{POP}). Knowledge has an exact characterisation in constructor theory: it is information that can act as a constructor and cause itself to remain instantiated in physical substrates. 

Crucially, error-correcting the replication is necessary. Hence the subunits $p_i$ must assume values in a {\sl discrete} ({digital}) information variable: one whose attributes are separated by non-allowed attributes. For, if all values in a continuum were allowed, error-correction would be logically impossible.

\subsubsection{Appearance of design} \label{DES}
 
Something with the appearance of design is often described as \qq{improbable} \cite{DAWBLIND, DAWIMPRO}. This is misleading because probability measures are multiplicative; so that would mean that two independent objects with the appearance of design would have much more of that appearance than they do separately. But that is not the case when the two objects have unrelated functionalities (such as, say, internal organs of different organisms). In contrast, two organs {\sl in the context of the same} organism, coordinating to the effect of gene propagation, do have a greater appearance of design than either separately. This can be expressed naturally in constructor-theoretic terms for programmable constructors. 

Consider a recipe $R$ for a possible task $T$. 
A sub-recipe $R'$ for the task $T'$ is {\bf fine-tuned} to perform $T$ if almost any slight change in $T'$ would cause $T$ to be performed to a much lower accuracy. (For instance, changing the mechanism of insulin production in the pancreas even slightly, would impair the overall task the organism performs.)
A programmable constructor $V$ whose repertoire includes $T$  {\sl has the appearance of design} if it can execute a recipe for $T$ with a hierarchical structure including {\sl several, different} sub-recipes, {\sl fine-tuned} to perform $T$. 
Each fine-tuned sub-recipe is performed by a sub-constructor contained in $V$: the number of fine-tuned sub-recipes performable by $V$ is a measure of $V$'s appearance of design. This constructor-theoretic definition is non-multiplicative, as desired.
%
\subsection{The logic of self-reproduction}\label{SEC:SR}

I shall now apply the results of section \ref{SEC:CAR} to self-reproduction, to conclude that no-design laws permit an accurate self-reproducer, provided that it operates via what I call, adapting Dawkins' terminology \cite{DAWREVE}, the {\bf replicator-vehicle logic}.

A self-reproducer $S$ (of the kind \eqref{REP2}) is a constructor for its own construction, from generic resources only. From the argument in \ref{SEC:CAR}  it follows that for $S$ to be a good approximation to a constructor for another $S$, it must consist of: a modular {\bf replicator}, $R=(r_1,r_2,\dots,r_n)$, instantiating the recipe for $S$ (the elementary units $r_i$ have attributes in an information variable $\Sigma$, corresponding to instructions); a programmable constructor, the {\bf vehicle} $V$, executing the recipe {\sl blindly}, i.e., implementing non-specific sub-tasks. 

The recipe instantiated by the replicator $R$ must contain all the knowledge about how to construct $S$, specifying a procedure for its construction. Note, however, that the recipe is in one sense incomplete:  as remarked in section \ref{SEC:CAR}, the recipe is not required to include instructions for the elementary tasks, which occur spontaneously in nature. These are indeed relied upon during actual cell development - they constitute epigenetics and environmental context. As remarked by George C. Williams, \qq{Organisms, wherever possible, delegate jobs to useful spontaneous processes, much as a builder may temporarily let gravity hold things in place and let the wind disperse paint fumes}, \cite{WIL2}.  

Under no-design laws, maintenance and error-correction are necessary for a high and improvable accuracy to be achieved; and in self-reproduction, crucially, it must be $S$ {\sl only} that brings about the new instance of $S$.  Therefore, since the maintenance cannot be performed by the laws of physics either, because of the no-design conditions, it must be executed by $S$. As in the general case of section \ref{SEC:CAR}, maintenance must be achieved via {\bf copying} the recipe and {\bf constructing} the vehicle $V$. These are enacted, respectively, by two sub-constructors in the vehicle, $C$ and $B$, which implement the replicator-vehicle logic that von Neumann discovered, \cite{VONBU}. 

In the {\bf construction phase} $B$ executes $R$ to construct a new vehicle $V$: 
\begin{equation}
N\overset{\rm B[R]} {\implies} (V,W)\;.\nonumber
\end{equation}
In bacteria $B$ includes the mechanisms for constructing the daughter cell, such as the ribosome which uses DNA instructions (translated into RNA) to construct proteins. Blind error-correction is possible via checks on the subtasks of the recipe; however, construction errors are not propagated, because the new vehicle is the result of executing the recipe in the replicator, {\sl not} a copy of the former vehicle.  

In the {\bf copy phase}, the blind replication of $R$ is performed by $C$, a copier of the information variable $\Sigma$:
\begin{equation}
(R,N)\overset{\rm C} {\implies}(R, R,W) \;.\label{COPY}
\end{equation}
This happens by replicating the {\sl configuration} of $R$ {\sl blindly}, one elementary unit at a time. It follows that $C$ is a universal copier for the set of replicators consisting of elementary units drawn from $\Sigma$ (a property called {\sl heredity} \cite{SZA2}).  Error-correction can happen blindly too, for instance via mismatch-repair. In bacteria this phase is DNA replication and $C$ includes all the relevant enzymes in the cell. \footnote{I do not model details irrelevant to the self-reproduction logic (e.g. DNA semi-conservative replication).  }
For the two phases to perform maintenance, the recipe for the vehicle $V$, instantiated in the replicator $R$, must be copied in the copy-phase. This requires the elementary instructions of the recipe to be (sets of) the elementary units $r_i$ of the replicator. In bacteria they are the codons - triplets of the elementary units of the replicator (the nucleotides), coding for the building blocks of proteins (aminoacids). 

The replication of each sub-unit $r_i$ constitutes a {\sl measurement} of which attribute $r_i$ holds, followed by constructing a new instance of it. Since the replicator $R$ must contain all the knowledge about $S$, the attributes in $\Sigma$, of which $R$ is made, must be {\sl generic resources}, so as to require no recipe (other than $R$) to be constructed from generic resources. I call a modular replicator such as $R$ whose subunits are made of generic resources a {\bf template replicator}. 
A DNA strand is one: the information variable $\Sigma$ is the set of nucleotides - they are simple enough to have been naturally occurring in pre-biological environments. 

We thus see that the two maintenance phases achieve self-reproduction, as they amount to bringing about a new $R$, by copying the former $R$, and a new $V$, by construction - controlled by $R$. Thus,  self-reproduction is stable precisely because copying and construction automatically execute the maintenance of $S$, by replicating the recipe and re-constructing the vehicle before the former instance of $S$ wears out; and they permit error-correction. 
For arbitrarily high accuracy, both phases implement elementary sub-recipes that are non-specific to self-reproducers, and do not bear design. Therefore arbitrarily accurate self-reproduction is permitted by no-design laws, provided that the latter allow replicators - i.e., information media.

\medskip

Rewriting the copy phase, \eqref{COPY}, as $$N\overset{\rm C[R]} {\implies} (R,W)\;,$$ to highlight that $C$ executes $R$, we see that a template replicator has a special property. It instantiates {\sl a recipe for its own construction from generic resources only} ($C$ does not need to contain any additional recipe to construct the subunits of $R$: it blindly copies the pattern, subunit by subunit; and the units are generic resources). This is unique to template replicators. No other object could be a recipe for the construction of itself to a high accuracy. For the argument in section \ref{SEC:CAR} implies that an instance (or a blueprint) of an object is not, in general, a recipe for its construction from generic substrates. A 3-D raster-scanner provided with an instance of, say, a bacterium could not re-produce it accurately from generic substrates {\sl only}: without a recipe containing the knowledge about the bacterium's structure, there would be no criterion for error-correction, resulting in a bound on the achievable accuracy. Likewise, an entire organism could not self-reproduce to a high accuracy via self-copying: without the recipe informing error-correction, an \qq{error catastrophe} \cite{RID} would occur.  This also provides a unifying descriptions of the two phases: the replicator $R$ is a recipe for another instance of itself, when instructing $C$; a recipe for the construction of another vehicle, when instructing $B$. Overall, it instantiates {\sl the full recipe for $S$} - see the figure \ref{SR}.

\begin{figure}[h]
\centering
\includegraphics[scale=0.4]{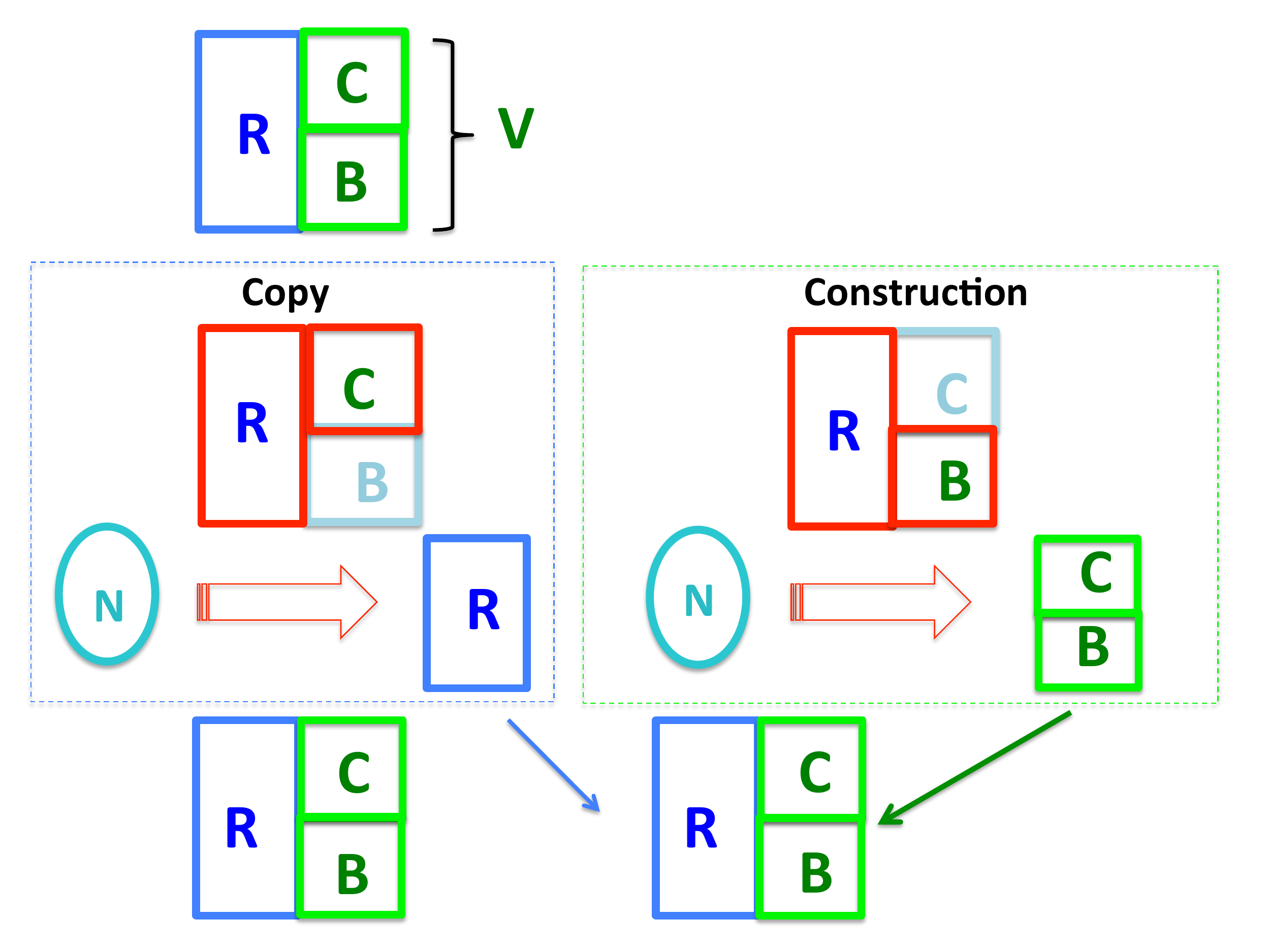} 
\caption{{\bf The logic of self-reproduction} 
{\small An accurate self-reproducer (top of the figure) consists of the replicator $R$ (blue outline) and the vehicle $V$ (green outline) - which contains the copier $C$ and the constructor $B$. In the copy phase $C$ copies the replicator $R$ - $C[R]$ (red outline) acts as a constructor. In the construction phase $B$ executes the recipe in $R$ to build a vehicle from generic resources $N$ - $B[R]$ (red outline) acts as a constructor. 
Finally (bottom) the copy of $R$ and the newly constructed vehicle form the offspring. }}
\end{figure}\label{SR}

$R$ is an {\sl active}, {\sl germ line} replicator \cite{DAWREVE}, because instantiates all the knowledge {\sl necessary to achieve its own replication}. It is a consequence of the above argument that high-fidelity replication is possible under no-design laws too, provided that there is a vehicle that performs blind copying and error-correction.  Moreover, for the replicator to preserve its ability to be an accurate replicator across generations, its vehicle must be reproduced too - together, they must constitute a self-reproducer. Hence self-reproduction is essential to high-fidelity replication under no-design laws. 

\subsection{Natural selection is permitted under no-design laws}\label{NAT}

These conclusions imply that an accurate self-reproducer - together with an accurate replicator - is permitted under no-design laws that allow for information media. So, under such laws, it can be constructed from generic resources only, given {\sl enough knowledge}: it could continue to exist, say, had a chemical lab created it.

However, one must also address the question: {\sl can} accurate self-reproducers arise from generic resources only, under such laws?

Note that what the prevailing conception would aim to prove is that the emergence of accurate self-reproducers {\sl follows} (with some probability) given certain initial conditions and laws of motion. This approach, informing the search for viable models for the origin of life, \cite{WADA},  is suitable to solve scientific problems such as predicting the existence of life elsewhere in the universe - e.g., by providing bounds to how probable the emergence of those self-reproducers is on an earth-like planet. Here I am addressing a different problem: whether accurate self-reproducers are {\sl possible} under no-design laws. This is a theoretical (indeed, constructor-theoretic) question and can be addressed without resorting to predictions.  Indeed, the theory of evolution provides a positive answer to that question, provided that two further points are established. I shall argue for them in what follows. 

{\sl The first point} is that {\sl the logic of evolution by natural selection is compatible with no-design} laws because - in short - selection and variation are {\sl non-specific} to its end products. This can be seen by modeling the logic of natural selection as an approximate {\sl construction}, whose substrates are populations of replicators and whose (highly approximate) constructor is the environment. This occurs over a much longer time-scale than that of self-reproduction, whereby replicators - constructors on the shorter scale - become now substrates. 


Evolution relies upon populations being changed by variation and selection over the time-scale spanning many generations.
Crucially, the mutations {\sl in the replicators}, caused by the environment, are {\sl non-specific}, (as in section \ref{SEC:CAR}), to the \qq{end product} of evolution (as Dawkins put it, not \qq{systematically directed to improvement} \cite{DAWBLIND}). This constructor-theoretic characterisation of mutations  replaces the less precise locution \qq{random mutations} (as opposed to non-random selection, \cite{DAWSELF}). 
These mutations are {\sl all} transmitted to the successfully created individuals of the next generation, by heredity - irrespective of their being harmful, neutral  or beneficial in that particular environment.

Selection emerges from the interaction between the replicators and the environment with {\sl finite} resources. It {\sl may} lead to equilibrium, given enough time and energy. If so, the surviving replicators are near a local maximum of effectiveness at being replicated in that environment. 

Thus, the environment is {\sl passive} and {\sl blind} in this selection process. Since it retains its ability to cause non-specific variation and passive selection again, it qualifies as a naturally-occuring { approximation} to a constructor. Crucially, it is a crude approximation to a constructor: crude enough that it could have arisen by chance and requires no explanation. Its actions - variations and selection - require no design in laws of physics, as they proceed by non-specific, elementary steps. So the logic of evolution by natural selection is compatible with no-design laws of physics.

{\sl The second point} is that natural selection, {\sl to get started}, does not require accurate self-reproducers with high-fidelity replicators. Indeed, the minimal requirement for natural selection is that each kind of replicator produce {\sl at least one viable offspring}, on average, per lifetime - so that the different kinds of replicators last long enough to be \qq{selected} by the environment. 
In challenging environments, a vehicle with many functionalities is needed to meet this requirement. But in unchallenging ones  (i.e. sufficiently unchanging and resource-rich), the requirement is easily met by highly inaccurate self-reproducers that not only have no appearance of design, but are so inaccurate that they {\sl can} have arisen spontaneously from generic resources under no-design laws - as proposed, for instance, by the current theories of the origin of life \cite{MAZA, SZA1}. For example, template replicators, such as short RNA strands \cite{SZA2}, or similar \qq{naked} replicators (replicating {\sl with poor copying fidelity} without a vehicle) would suffice to get natural selection started. Since they bear no design, they require no further explanation -  any more than simple inorganic catalysts do.\footnote{Metabolism-first conjectures such as Dyson's, \cite{DYS}, based on evolution of molecules \qq{by random drift}, without template replicators, are thus bound by the present argument to rely on elementary operations, compatible with no-design requirements. }

I conclude that the theory of evolution is compatible with no-design laws of physics, that allow, in addition to enough time and energy, information media.  These requirements do not contain the design of biological adaptations. Hence, under such laws, the theory of evolution fully explains the appearance of design in living organisms, without their being intentionally designed.

\section{Self-reproduction under quantum theory}

In this section I shall show that accurate self-reproduction is compatible with quantum theory: after a critique of works claiming the opposite (section \ref{IRR}), I demonstrate a quantum-mechanical (kinematical) model of the replicator-vehicle logic (section \ref{QU}). 

\subsection{Irrelevance of the incompatibility arguments}\label{IRR}

The first claim that self-reproduction is incompatible with quantum physics was made by Wigner \cite{WIG}. Its agenda is to show that \qq{the present laws of quantum mechanics will have to undergo modifications before they can be applied to the problems of life} and they need to be complemented by \qq{biotonic} laws, containing the design of self-reproducers \cite{WIG}. The proposed method to do that is showing that the unitary laws of quantum physics which cause arbitrarily accurate self-reproduction of an organism $S$ constitute a vanishingly small fraction of all possible unitaries, {\sl when $S$ is a sufficiently specialised entity} (has the appearance of design).

In Wigner's model the substrates of self-reproduction consist of three parts: the parent self-reproducer; the substrates to be transformed into the new instance and the substrates to be transformed into waste. Correspondingly, the total Hilbert is modelled as $\tilde {\cal H}={\cal H}_1\otimes {\cal H}_2 \otimes {\cal H}_3$, where the labels 1,2,3 refer, respectively, to those three parts, and ${\cal H}_1\sim {\cal H}_2$ (I shall denote both by ${\cal H}$). The \qq{highly specialized} self-reproducer is a subspace $W_S\subset {\cal H}$ whose dimension $h(S)$ is much {\sl smaller} than the dimension $d$ of ${\cal H}$.  
Wigner's argument shows that the set of unitaries causing the replication of $W_S$ in a given tensor-product structure \begin{equation}\{U:\;\exists W_S\subset {\cal H},\;\exists \ket{N}\in {\cal H}\otimes {\cal H}_3\;:\; W_S\otimes {\rm Span}\{\ket{N}\} \rightarrow W_S\otimes W_S\otimes {\cal H}_3\}\nonumber \end{equation} has measure which tends to zero as ${h(S)\over d}\rightarrow 0$ (with respect to the natural measure on the space of unitaries) \cite{BAE}.  

Wigner concludes that unless the unitary $U$ is \qq{tailored so as to permit self-reproduction, it is infinitely unlikely} that, under quantum theory, accurate self-reproduction of specialised entities can occur; whence the need for designed laws.

Evidently, this argument would not rule out self-reproduction only. It would apply to all the unitaries $U$ : $W_C\otimes {\rm Span}\{\ket{N}\} \rightarrow W_C\otimes W_T\otimes {\cal H}_3$ for some subspaces $W_C$, $W_T$ whose dimension is smaller than $d$. Hence it would rule out, under Wigner's interpretation, {\bf every} specialised construction. 

But the interpretation is erroneous. As explained, the  \qq{non-typical}interaction is compatible with no-design laws (and in particular with quantum mechanics, see section \ref{QU}), because it can be decomposed into elementary steps - non-specific to $S$ - controlled by the recipe.  No-design laws plus a knowledge-laden recipe can play the role that Wigner erroneously assumed can only be played by knowledge-laden laws and a generic initial state. Also,  the \qq{difficult feat} \cite{WIG} of bringing about the knowledge in the recipe does not require intentional design, as explained by  evolutionary theory, which I showed is compatible with no-design laws of physics. 

The misconception underlying Wigner's interpretation is to identify the mathematical property of being a \qq{non-typical} unitary with the {\sl physical} property of containing the design of an object. Evidently the former does not imply the latter; so, the argument is irrelevant to the claim. Similarly, the (multiplicative) property of belonging to a small subspace misrepresents the appearance of design (which is non-multiplicative, see section \ref{DES}). 

Moreover, as pointed out in \cite{BAE}, Wigner's {\sl argument} is about an over-con\-stra\-i\-ned set of unitaries, i.e., the ones causing reproduction of $W_S$ in a tensor-product structure that is fixed {\sl a priori}. But Wigner\aa s purpose is served by the set of unitaries with the property that {\sl there exists} a tensor product structure in which they would cause self-reproduction. Nevertheless, Baez's theorem, \cite{BAE}, that almost all unitaries would achieve replication of a {\sl single state} in the presence of a {\sl specific} initial state, in {\sl some} tensor product structure, is not actually a rebuttal of Wigner\aa s {\sl claim}. One could reach the same conclusion as Wigner's by arguing that this initial condition is in fact of zero-measure in the set of all possible initial conditions. 
Also, the replication of a single quantum state (which Wigner also discusses) is too strict a requirement to model self-reproduction of living entities, as it does not permit evolution.


Confusing self-reproduction and replication (cloning) of single quantum states has informed another claim, that self-reproduction of a universal constructor with  finite resources is forbidden by quantum theory \cite{BRA, CHAPRA, PATI}.  The model supporting this claim comprises a collection of substrates, with Hilbert space ${\cal K}$, $n$ of which are the raw materials, $\ket{0}\in{\cal K}$; the rest contains the processor, the control unit and the program space of the alleged universal constructor. $\ket{\psi}\in {\cal K}^{m-1} $ is any state of the processor and $\ket{P_{\psi}}\in {\cal K} : \ket{P_{\psi}}\ket{0}\rightarrow \ket{P_{\psi}}\ket{\psi}$ is the program for the state $\ket{\psi}$.  Self-reproduction of the universal constructor would correspond to a unitary satisfying, for some states $\ket{C}$, $\ket{C^*}$ of the control unit: $$L(\ket{\psi}\ket{P_{\psi}}\ket{0}^{\otimes m}\ket{C})\ket{0}^{\otimes n -m} = \ket{\psi}\ket{P_{\psi}}L(\ket{\psi}\ket{P_{\psi}}\ket{0}^{\otimes m}\ket{C^*})\ket{0}^{\otimes n - 2m} \;,$$ for {\sl every} ${\psi}$. This is impossible, the argument goes, unless programs for different states are orthogonal; in which case (allegedly) infinitely many resources would be needed, as the program space would have to be infinite-dimensional. 

This claim, too, is irrelevant to whether living self-reproducers are compatible with quantum mechanics. $L$ copies  {\sl each state} of the vehicle and the program for that state, while actual self-reproduction requires the re-production of a {\sl subspace} - the property of being a self-reproducer. Indeed, $L$ is ruled out by the no-cloning theorem, if the programs are not orthogonal. Besides, actual self-reproducers need not be  universal constructors: their repertoire need only include (and, in the earth's biosphere, does include) very few products, compared with the set of all possible products.
 
\subsection{The replicator-vehicle logic under quantum theory}\label{QU}

I shall now demonstrate a quantum-mechanical model of self-reproduction, implementing the replicator-vehicle logic. I model the world as a collection of replicas of the substrates that can have the attribute of being a self-reproducer. Each substrate has Hilbert space ${\cal H}={\cal H}_r\otimes {\cal H}_v$, where ${\cal H}_r $ is the space of the replicator and ${\cal H}_v$ that of the vehicle. One replica contains the parent, one its offspring, and the remaining $w$ are transformed into waste products. The law of motion is a unitary $U$ which I shall prove to be compatible with self-reproduction. 
The attribute $A$ is the $+1$-eigenspace of the projector $\hat A$ for holding that attribute: $A=\{\ket{\psi}\;:\;\hat A\ket{\psi}=\ket{\psi}\}$.

Let $\hat N={  {\hat N_r}}\otimes { {\hat N}_v}$ (defined on ${\cal H}_r\otimes {\cal H}_v)$ be the projector for being generic substrates and $\hat S={  {\hat R_s}}\otimes { {\hat V}_s}\;$ be that for being a self-reproducer S, where ${\hat V}_s$ is the projector (defined on ${\cal H}_v$) for being a vehicle and ${\hat R_s}$ is the projector (defined on  $H_r$) for being a recipe for it. 

For evolution to be possible a set $\Sigma$ of different self-reproducers must be allowed in the environment $N$. So, the unitary law of motion $U$ must satisfy, $\forall s,s'\in \Sigma$:  

\begin{equation}
U\;:\;{R_{s'}}\otimes {{V}_s}\otimes   N^{\otimes (w+1)}\rightarrow   {R_{s'}}\otimes {{V}_s}\otimes   {R_{s'}}\otimes {V}_{s'}\otimes   {\cal H} ^{\otimes w}\;.\label{repl-Q}
\end{equation}
This is self-reproduction as in equation \eqref{REP2} (for $s=s'$); and each self-reproducer has heredity: a vehicle copies any recipe (coding for vehicles or not); and executes recipes to construct any other vehicle. Different vehicles are represented by mutually orthogonal projectors. By unitarity of \eqref{repl-Q}, recipes for different self-reproducers are orthogonal too: $\hat R_s  \hat R_{s'}=0$. I shall confine attention to a basis of orthogonal programs spanning each subspace (their superpositions code, by linearity, for the same vehicle). 

That the environment contains no knowledge about the self-reproducer is guaranteed by imposing the (sufficient) conditions $\hat R_s\hat N_r=0$, $\hat V_s \hat N_v=0$, for all $s$. 

Let $h_{\hat A}$ be the rank of the projector $\hat A$, $d_r$ the dimension of ${\cal H}_r$, $d_v$ that of ${\cal H}_v$. Each self-reproducer occupies a small volume of the space of all possible states of a system, whereby $h_{\hat R_s} \ll d_r$, $h_{\hat  V_s}\ll d_v$.  So do generic resources, being a collection of low-entropy, low-entanglement states: therefore, $h_{ \hat  N_r} \ll d_r$, $h_{\hat N_v} \ll d_v$. Thus the condition in \eqref{repl-Q} can be met by many unitaries, because $$h_{\hat R_{s'}}h_{\hat V_s}(h_{\hat N_r}h_{\hat N_v})^{w+1}\ll h_{\hat R_{s'}}^2h_{\hat V_s}h_{\hat V_{s'}}(d_rd_v)^{w}\;, \;\;\forall s,s'\in \Sigma\;.$$

Each unitary permitting self-reproduction is the serial composition of one implementing the {\sl copy phase} and one implementing the {\sl construction phase}. Without loss of generality, I adopt a qubit model, where the replicator consists of $r$ qubits and the vehicle of $v$ qubits. The information variable $\Sigma_b$ representing the elementary replicator units comprises, say, the $z$-component eigenvectors of a single qubit. 
I model generic resources as a fixed input state from $N$, say the simultaneous $+1$-eigenvector of the $z$-components of the qubits in ${\cal H}^{w+1}$ (having the desired features of low-entropy and no-entanglement). 

The {\bf copy} unitary ${\bf C}$ (replicating the recipe) must satisfy:  
\begin{eqnarray}
{\bf C}\;:\;{R_{s'}}\otimes N_r\otimes {V}_s\otimes   N_r^{\otimes w}\rightarrow   {R_{s'}}\otimes   {R_{s'}}\otimes { {V}_s}\otimes {\cal H } _r^{\otimes w}\;\;,\;\forall s,s'\in \Sigma\;\nonumber 
\end{eqnarray}
This is realised as a sequence of ${\small CNOT}_{i,j}$ from qubit $i$ in the parent's replicator subspace ${\cal H}_r$ to the qubit $j$ of the new instance's, performed in the presence of any vehicle:
$$
{\bf C}=\prod\limits_{i=1}^{r}\displaystyle{{\small CNOT}_{i, i+r}}\otimes   \sum_s  {V_s}+ W_c\otimes ({\rm id}-  \sum_s  {V_s})
$$ 
where the unitary $W_c$ occurs if no vehicle is available. (A possible error-correction process, controlled by the program, is not modeled here.)

The {\bf construction} unitary ${\bf B}$ must satisfy: 
\begin{equation}
{\bf B}\;:\;{R_{s'}}\otimes { {V}_s}\otimes   N_v^{\otimes (w+1)}\rightarrow   {R_{s'}}\otimes { {V}_s}\otimes   {V_{s'}}\otimes {\cal H }_v ^{\otimes w}\;\;,\;\forall s,s'\in \Sigma\;\nonumber
\end{equation}
which can be realised by the unitary 
$$ {\bf B}= \sum\limits_{s\in \Sigma}  R_s\otimes U_{s}+ \left ({\rm id}-\sum\limits_{s\in \Sigma} {R}_{s}\right)\otimes W_b  \;.
$$ 
Here, any vehicle executes the recipe $R_s$ to construct a new vehicle $V_s$, via $U_s= \sum_s  {V_s}\otimes B_{s}+({\rm id}-  \sum_s  {V_s})\otimes \tilde W_b\;$ where the unitary $B_{s}$ is the construction of a new instance of the vehicle (possibly including error-correction): $B_{s}\;:\;  N_v^{\otimes (w+1)} \rightarrow V_s\otimes {\cal H}_v^{\otimes w}\:.$ The arbitrary unitaries $\tilde W_b$ and $ W_b$ express, respectively, the output in the absence of a vehicle and of a program coding for a vehicle. 

Under quantum theory $U_{s}$ is decomposable into elementary operations, conditioned on groups of qubits in the replicator: it is controlled by the recipe only. It can be implemented without the design of $V_s$ being encoded in the underlying laws. In addition, (by universality) a decomposition into elementary (coherent) quantum gates is allowed. Whether quantum coherence could actually be used, either in living or in artificial self-reproducers, e.g. to enhance the construction efficiency, is an open question in quantum biology, \cite{QBIO}. But I have just shown that this possibility is {\sl allowed}. Hence self-reproduction is compatible with quantum theory: it can be implemented and stabilised via elementary operations requiring no design, as promised. 

\section{Conclusion}

I have proved that the physical processes the theory of evolution relies upon are possible under no-design laws, provided that the latter allow for information media (and enough generic resources). Under such laws, accurate self-reproduction can occur, but only via von Neumann's replicator-vehicle logic; and a high fidelity replicator requires an accurate self-reproducer. My argument also highlights that all accurate constructors, under such laws, must contain {\sl knowledge} - a special abstract constructor - in the form of a recipe, instantiated in a replicator.  

I have also extended von Neumann's model of the {\sl logic} of self-reproduction to quantum theory. This informs further investigations of quantum effects in natural and artificial self-reproducers.
Constructor theory has also expressed exactly within fundamental physics, the logic of self-reproduction, replication, and natural selection, and the appearance of design. This has promise for a deep unification in our understanding of life and physics. 

\newpage 

\section*{\small{Acknowledgements}}\nonumber

\small{I thank Simon Benjamin and Mario Rasetti for many helpful comments; Paul Davies and Sara Walker for deep conversations on replicators and early life; Alan Grafen for illuminating conversations on neo-Darwinism; and especially David Deutsch for fruitful discussions and incisive criticism on earlier drafts of this paper. I also thank three anonymous referees for suggesting many significant improvements. This research is supported by the Templeton World Charity Foundation. }

 \end{document}